\documentclass[letterpaper]{article}
\usepackage{aaai}

\usepackage{amsmath}
\usepackage{amssymb}
\usepackage{times}
\usepackage{helvet}
\usepackage{courier}
\usepackage{subfigure}
\usepackage{graphicx}
\usepackage{epstopdf}
\usepackage{url}
\newcommand{\Prob}{\text{Prob}}

\newcommand{\qed}{\nobreak \ifvmode \relax \else
      \ifdim\lastskip<1.5em \hskip-\lastskip
      \hskip1.5em plus0em minus0.5em \fi \nobreak
      \vrule height0.75em width0.5em depth0.25em\fi}

\title{Composite Social Network for Predicting Mobile Apps Installation}
\author{Wei Pan \and Nadav Aharony \and Alex (Sandy) Pentland\\
MIT Media Laboratory \\
20 Ames Street \\
Cambridge, Massachusetts 02139 \\
}
\begin{document} 
\maketitle
\begin{abstract}
\begin{quote}
We have carefully instrumented
a large portion of the population living in a university graduate dormitory 
by giving participants Android smart phones running our sensing software.
In this paper, we propose the novel problem of predicting mobile application 
(known as ``apps'')
 installation using social networks and explain its challenge. 
Modern smart phones, like the ones used in our study, are able to
collect different social networks using built-in sensors. (e.g. Bluetooth proximity network,
call log network, etc) While this information is accessible to app market
makers such as the iPhone AppStore, 
it has not yet been studied how app market makers can use these information
for marketing research and strategy development. We develop
a simple computational model to better predict app installation by using
a composite network computed from the different networks sensed by phones.
Our model also captures individual variance and exogenous factors in 
app adoption. We show the importance of considering all these
factors in predicting app installations, and we observe the 
surprising result that app installation is
indeed predictable. We also show that our model achieves the 
best results compared with generic approaches: our 
results are four times better than random guess, and predict
almost $45\%$ of all apps users install with almost $45\%$ precision ($F_1$ score$=0.43$).  
\end{quote}
\end{abstract}

\section{Introduction}
\noindent Recent research projects have demonstrated that social networks
correlate with individual behaviors, such as obesity~\cite{christakis2007spread} and diseases~\cite{colizza2007modeling}, to name two. Many large-scale networks are analyzed, and this field is becoming increasing popular~\cite{eagle2010network}~\cite{leskovec2007dynamics}.

We are interested in studying the network-based prediction for
mobile applications (referred as ``apps'') installation, as the mobile application business is growing 
rapidly~\cite{idc2010}. The app market makers, such as iPhone AppStore and Android Market, run on almost all modern smart phones, and they have access to phone data and sensor data.
As a result, app market makers can infer different types of networks, such as 
the call log network and the bluetooth proximity network, from phone data. However, 
it remains an unknown yet important question whether these data can be used for app marketing.
Therefore, in this paper we address
the challenge of utilizing all different network data obtained from smart phones for 
app installation prediction.
% Modern smart phones are capable of sensing different networks. For instance,
% the call logs can be used to infer one social network; Bluetooth radio on 
% phones also provides proximity network among individuals; Facebook data
% can also be used to infer friendship networks. While \emph{almost all other studies} 
% focus on one single type of social networks, in this study, our core idea
% is to infer a optimal composite network from all these candidate social networks to 
% maximize the prediction performance of app installation. We believe that
% since smart phones are increasingly popular these days, and app marketing companies
% should leverage all possible network information.

It is natural to speculate that there are network effects in users' app installation, but we eventually realize that 
it was very difficult to adopt existing tools from
large-scale social network research to model and 
predict the installation of certain mobile apps for each user due to the following facts:
\begin{enumerate}
\item The underlying network is not observable. While many projects
assume phone call logs are true social/friendship networks~\cite{zhang2010discovery}, 
others may use whatever network that is available as the
 underlying social network.
Researchers have discovered that call network may not be a good approximation ~\cite{eagle2006reality}.
On the other hand, smart phones can easily sense multiple networks using
built-in sensors and software: a) The
call logs can be used to form phone call networks; b) Bluetooth radio
can be used to infer proximity networks~\cite{eagle2006reality}; 
c) GPS data can be used to infer user moving patterns, and furthermore their
working places and affiliations~\cite{farrahi2010probabilistic}; d) Social network tools (such as 
the Facebook app and the Twitter app) can observe users' online
friendship network. In this work, our key idea is to infer an \emph{optimal composite network}, 
the network that best describes app installation, 
from multiple layers of different networks easily observed by modern
smart phones, rather than assuming a certain network as the real social network
explaining app installation.

\item Analysis for epidemics~\cite{ganesh2005effect} and Twitter networks~\cite{yang2010modeling} is 
based on the fact that network is the only mechanism for adoption. The only
way to get the flu is to catch the flu from someone else, and the only way to
retweet is to see the tweet message from someone else.
For mobile app, this is, however, not true at all.
Any user can simply open the AppStore (on iPhones) or the Android Market 
(on Android phones), browse over different lists of apps, and pick the one
that appears most interesting to the user to install without peer influence. 
One big challenge, which makes modeling the spreading of apps
difficult, is that one can install an app without any external influence
and information. One major contribution of this paper is that
we demonstrate it is still possible to build a tool to observe 
network effects with such randomness.

\item The individual behavioral variance in app installation is so significant that 
any network effect might possibly be rendered unobservable from the data.
For instance, some geek users may try and install all hot apps on the market, while many
inexperienced users find it troublesome even to go through the process
of installing an app, and as a result they only install very few apps.

\item There are exogenous factors in the app installation behaviors. 
One particular factor is the popularity of apps. For
instance, the Pandora Radio app is vastly popular and highly ranked in the
app store, while most other apps are not. Our model takes this issue into account too, 
and we show that exogenous factors are important in increasing prediction precision.
\end{enumerate}

Classic diffusion models such as Granovetter's work~\cite{granovetter1983threshold} 
are applicable to simulation, but lack data fitting and prediction
powers. Statistical analysis used by social scientists
such as matched sample estimation ~\cite{aral2009distinguishing} are
only for identifying network effects and mechanism. Recently works
in computer science for inferring network structure assume simple
diffusion mechanism, and are only applicable to artificial simulation data
on real networks~\cite{gomez2010inferring}~\cite{myers9convexity}. 
On the other hand, our work addresses the above issues in practical
 app marketing prediction. On the mobile-based behavioral prediction side, 
The closest research is the churn prediction problem 
in mobile networks~\cite{richter2010predicting},
which uses call logs to predict users' future decisions of switching
mobile providers. 
To our knowledge, we don't see other related works for similar problems.

\section{Data}
We collected our data from March to July 2010 with 55 participants, who are
residents living in a married graduate student residency of a major
US university. Each participant is given an Android-based cell phone 
with a built-in sensing software developed by us. The software runs 
in a passive manner, and it didn't interfere the normal usage of the phone.

Our software is able to capture all call logs in the experiment
period. We therefore obtained a call log network between all participants
by treating participants as nodes and the number of calls between two nodes
as weights for the edge in-between.
The software also scans near-by phones and other Bluetooth devices every
five minutes to capture the proximity network between individuals.
The counts on the number of Bluetooth hits are used as edge weights 
similar to the call log network as done in Eagle et al~\cite{eagle2006reality}.
We have also collected the affiliation network and the
friendship network by deploying a survey,
which lists all the participants and ask each one to list their affiliations (i.e. the academic department), and rate their relationships with everyone else
in the study. We believe for app market makers the affiliation network can also be inferred simply
by using phone GPS/cell tower information as shown by Farrahi et al\cite{farrahi2010probabilistic}.
However, this is not the focus of this work, and here we simply 
use survey data instead.
Though the friendship network is also
collected using surveys, we suggest that the app market makers 
can obtain the friendship network 
from phones by collecting data from social networking apps such as 
the Facebook and Twitter apps.
We summarize all the networks obtained from both phones and surveys
in Table ~\ref{table1}. We refer to all networks in Table ~\ref{table1}
as \emph{candidate networks}, and all candidate networks will be used
to compute the optimal composite network. 
It should be noted that all networks are reciprocal in this work.

We want to emphasize the fact that the network data we used in Table \ref{table1}
are obtainable for app market makers such as Apple iTunes Store, as 
they have access to phone sensors as well as user accounts. 
Therefore, our approach in this paper can be beneficial to them for marketing
research, customized app recommendation and marketing strategy making.

\begin{table*}
\centering
\begin{tabular}{ |c|c| c| c| } 
\hline 
Network & Type & Source & Notation \\
\hline 
Call Log Network& Undirected,Weighted & \# of Calls & $G^c$ \\
Bluetooth Proximity Network & Undirected,Weighted & \# of Bluetooth Scan Hits & $G^b$ \\
Friendship Network & Undirected,Binary & Survey Results (1: friend; 0: not friend)  & $G^f$ \\ 
Affiliation Network & Undirected,Binary & Survey Results (1: same; 0: different) & $G^a$ \\
\hline 
\end{tabular}
\caption{Network data used in this study. }
\label{table1}
\end{table*}

\begin{figure*}[ht]
\centering
\subfigure[]{
\label{fig1a}
\includegraphics[width=0.35\textwidth]{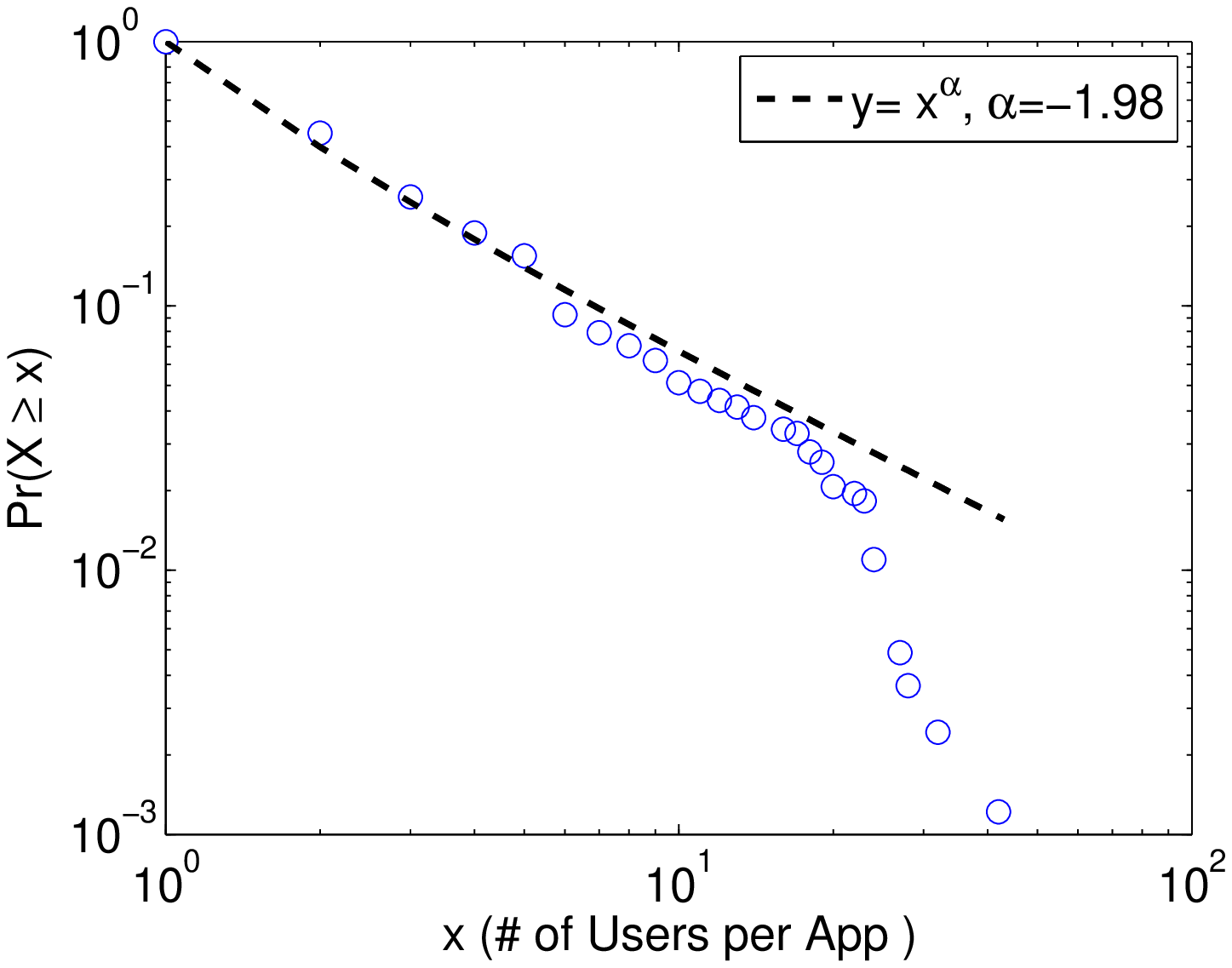}
} 
\subfigure[]{
\label{fig1b}
\includegraphics[width=0.35\textwidth]{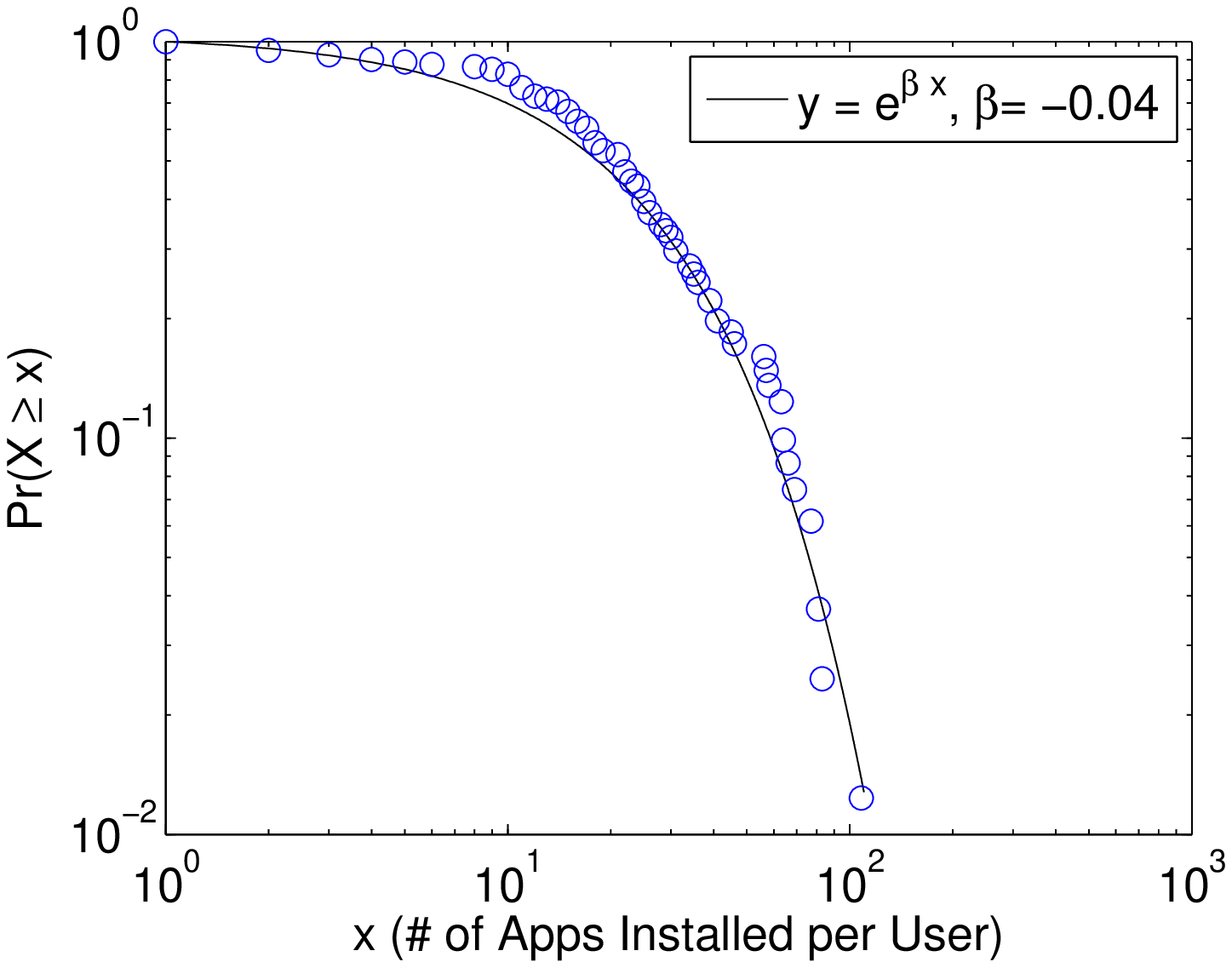}
}
\caption{Circles are real data, and lines are fitting curves. Left: Distribution of number of users for each app. Right: Distribution of number of apps each user installed. }
\end{figure*}

Our built-in sensing platform is constantly monitoring the installation of mobile
apps. Every time a new app is installed, this information will be collected
and sent back to our server within a day. Overall, we receive a total of 
821 apps installed by all 55 users. Among them, 173 apps have at least two
users. For this analysis, we only look at app installations and ignore 
un-installations. We first demonstrate statistics for all of the apps 
in the study: In Fig. ~\ref{fig1a}, we plot the distribution of 
number of users installing each app.
We discover that our data correspond very well with 
a power-law distribution with exponential cut.
In Fig. ~\ref{fig1b}, we plot the distribution of number of apps installed per user, 
which fits well with an exponential distribution.

Fig. \ref{fig1a} and \ref{fig1b} illustrate detailed insight into our dataset.
Even with a small portion of participants, the distribution characteristic is clearly observable. 
We find that apps have a power-law distribution of users, which suggests
that most apps in our study community have a very small user pool, and
very few apps have spread broadly. The
exponential decay in Fig. \ref{fig1b} suggests that the variance of individual
user is significant: There are users having more than 100 apps installed,
and there are users having only a couple of apps.

\section{Model}
In this section, we describe our novel model for capturing
the app installation behaviors in networks. In the following content,
${\mathbf G}$ denotes the adjacency matrix for graph $G$.
Each user is denoted by $u \in \{1,...,U\}$. Each app
is denoted by $a \in \{1,...,A\}$. We define the binary random
variable $x_u^a$ to represent the status of adoption (i.e. app installation): 
$x_u^a=1$
if $a$ is adopted by user $n$, $0$ if not. 

As introduced in the previous section, the
different social relationship networks that can be
inferred by phones are denoted by
${\mathbf G}^1,...,{\mathbf G}^M$. Our model
aims at inferring an optimal composite network ${\mathbf G}^{\text{opt}}$ with
the most predictive power from all the candidate social 
networks. The weight of edge $e_{i,j}$ in graph ${\mathbf G}^m$ is
denoted by $w^m_{i,j}$. The weight of an edge in
${\mathbf G}^{\text{opt}}$ is simply denoted by $w_{i,j}$.

\subsection{Adoption Mechanism}
One base idea of our model is the
non-negative accumulative assumption, which distinguishes 
our model from other linear mixture models.  We define ${\mathbf G}^{\text{opt}}$
to be:

\begin{equation}
{\mathbf G}^{\text{opt}} = \sum_m \alpha_m {\mathbf G}^m, \text{where } \forall m, \alpha_m \geq 0.
\label{optg} 
\end{equation}

The intuition behind this non-negative accumulative assumption
is as follows: if two nodes are connected by a certain type of
network, their app installation behaviors may or may not
correlate with each other; On the other hand, if two nodes are not connected
by a certain type of network, the absence of the link between
them should lead to \emph{neither positive or negative effect} 
on the correlation between their app installations. 
As shown in Table \ref{comp1} in the experiment session, 
our non-negative assumption
brings significant performance increase in prediction. Non-negative
assumption also makes the model stochastic and theoretically sound.
We treat binary graphs as weighted graphs as well. Since
$\alpha_1,...,\alpha_M$ is the non-negative weights for each candidate
network in describing the optimal composite network. 
We later refer to the vector $(\alpha_1,...,\alpha_M)$
as the optimal composite vector. Our non-negative
accumulative formulation is also similar to 
mixture matrix models in machine learning literature~\cite{el2008better}.

We continue to define the network potential $p_a(i)$:

\begin{equation}
p_a(i)=  \sum_{j \in {\mathcal N}(i)} w_{i,j} x_j^a,
\label{nb}
\end{equation}
where the neighbor of node $i$ is defined by:
\begin{equation}
{\mathcal N}(i) = \{j|\exists m \text{ s.t. } w^m_{i,j} \geq 0 \}.
\end{equation}

The potential $p_a(i)$ can also be decomposed into potentials
from different networks:

\begin{equation}
p_a(i) = \sum_m \alpha_m \underbrace{(\sum_{j \in {\mathcal N}(i)} w^m_{i,j} x_j^a)}_{p_a^m(i)},
\end{equation}
where $p^m_a(i)$ is the potential computed from one single candidate network.
We can think of $p_a(i)$ as the potential of $i$ installing
app $a$ based on the observations of its neighbors on the composite
network. The definition here is also similar to 
incoming influence from adopted peers for many cascade models~\cite{kempe2003maximizing}.

Finally our conditional probability is defined as:
\begin{equation}
\Prob(x^a_u=1 | x^a_{u'}: u'\in {\mathcal N}(u)) = 1 - \exp(-s_u - p_a(u)),
\label{prob} 
\end{equation}
where $\forall u, s_u \geq 0$. $s_u$ captures 
the individual susceptibility of apps, regardless
of which app. We use the exponential function for two reasons:
\begin{enumerate}
\item The monotonic and 
concave properties of $f(x)= 1-\exp(-x)$ matches with recent
research~\cite{centola2010spread}, which suggests that the probability 
of adoption increases at a decreasing rate with increasing external
network signals.
\item It forms a concave optimization problem during 
maximum likelihood estimation in model training. 
\end{enumerate}
As shown in the experiment section and based on our experiences,
this exponential model yields the best performance.

\subsection{Model Training}
We move on to discuss model training. During the training
phase, we want to estimate the optimal values for the $\alpha_1,...,\alpha_M$ and 
$s_1,...,s_U$. We formalize it as an optimization problem by maximizing 
the sum of all conditional likelihood.

Given all candidate networks, 
a training set composed of a subset of apps $\text{TRAIN} \subset \{1,...,A\} $, and $\{x^a_u:\forall a \in \text{TRAIN}, u \in \{1,...,U\} \}$,  we compute:
\begin{eqnarray}
\lefteqn{\arg\max_{s_1,...,s_U, \alpha_1,...,\alpha_M} f(s_1,...,s_U,\alpha_1,...,\alpha_M),} \nonumber \\
&\text{Subject to:}& \forall u, s_u \geq 0, \forall m, \alpha_m \geq 0 \label{optimization} 
\end{eqnarray}

where:
\begin{eqnarray}
\lefteqn{f(s_1,...,s_U,\alpha_1,...,\alpha_M)} \nonumber \\
&=&
\log \bigg[\prod_{a \in \text{TRAIN}} \prod_{u: x^a_u=1 } \Prob(x^a_u=1 |x^a_{u'}: u' \in {\mathcal N}(u))  \nonumber \\
 && \prod_{u: x^a_u=0} \big(1 - \Prob(x^a_u=1 | x^a_{u'}:u' \in {\mathcal N}(u)) \big) \bigg] \nonumber \\
&=& \sum_{a \in \text{TRAIN}} \left[ \sum_{u:x^a_u=1} \log(1 - \exp(-s_u - p_a(u)) \right. \nonumber  \\
&& \left. - \sum_{u:x^a_u=0} \left(s_u + p_a(u)\right) \right] \label{optimization1} \\
\end{eqnarray}

This is a concave optimization problem. 
Therefore, global optimal is guaranteed, and there exist efficient algorithms 
scalable to larger datasets ~\cite{boyd2004convex}.
% though we don't have them yet. 
We use a MATLAB built-in implementation here, and
it usually take a few seconds during optimization in our experiments.

Compared with works on inferring networks~\cite{gomez2010inferring}~\cite{myers9convexity}, 
our work is different as
we compute $G^{\text{opt}}$ from existing candidates networks. 
In addition, we don't need any additional regularization term or tuning
parameters in the optimization process. 

We emphasize that our algorithm doesn't distinguish 
the causality problem~\cite{aral2009distinguishing} in network effects:
i.e.,we don't attempt to understand the different reasons why network neighbors
have similar app installation behaviors. It can either be 
diffusion (i.e. my neighbor tells me), or homophily (i.e. network 
neighbors share same interests and 
personality). Instead, our focus is on prediction of app installation,
and we leave the causality problem as future work.

\subsection{Virtual Network for Exogenous Factors}
Obvious exogenous factors include the popularity and quality of an app. The popularity
and quality of an app will affect the ranking and review of the app in the AppStore/AppMarket,
and as a result higher/lower likelihood of adoption. We can model this by introducing a
virtual graph $G^{p}$, which can be easily plugged into our composite network framework.
$G^p$ is constructed by adding a virtual node 
$U+1$ and one edge $e_{U+1,u}$ for each actual user $u$. The corresponding weight
of each edge $w_{U+1, u}$ for computing $p_a(u)$ is $C^a$, 
where $C^a$ is a positive number describing
the popularity of an app. In our experiment, we use the number of installations
of the app in this experimental community as $C^a$. We have been looking at
other sources to obtain reliable estimates for $C^a$, 
but we found that the granularity from public sources to be unsatisfying.
In practice for app market makers, we argue that 
$C^a$ can be easily obtained accurately by counting app downloads and 
app ranks. 

The exogenous factors also increase accuracy in measuring network effects for a non-trivial reason: 
Considering a network of two nodes connected
by one edge, and both nodes installed an app. If this app is very popular, then the fact
that both nodes have this app may not imply a strong network effect. On the contrary, 
if this app is very uncommon, the fact that both nodes have this app implies a strong 
network effect. Therefore, introducing exogenous factors does help our algorithm better
calibrate network weights.

\section{Experiments}
Our algorithm predicts the probability of adoption (i.e. installing an
app) given its neighbor's adoption status. $p_i \in \left[0,1\right]$ 
denotes the predicted probability of installation,
while $x_i \in \{0,1\}$ denotes the actual outcome.  The most common prediction measure
is the Root Mean Square Error ($\text{RMSE}=\sqrt{\frac{1}{n} \sum_{i=1}^n(p_i - x_i)^2}$).
This measure is known to assess badly the prediction method's ability~\cite{goel2010prediction}. 
Since in our dataset most users have installed very few apps, a baseline
approach can simply predict the same small $p_i$ and still achieve very low RMSE.

For app marketing, the key objective is not to know the probability
prediction for each app installation, but to rank and identify a sub-group of individuals
who are more likely to appreciate and install certain apps compared with
average users. Therefore, we mainly adopt the approach in rank-aware measures
from information retrieval practices~\cite{manning2008introduction}. For each app, we rank the likelihood of adoption
computed by prediction algorithms, and study the following factors:
\begin{itemize}
\item[a)] Mean Precision at $k$ (MP-$k$): We select the top $k$ individuals with
highest likelihood of adoption as predicted adopters from our algorithms, and 
compute precision at $k$ ($\frac{\text{\# true adopters among } k \text{ predicted adopters}}{k}$). 
We average precisions at $k$ among all apps in the testing set to get MP-$k$.  
On average each app has five users in our dataset. Therefore, the default value for $k$
is five in the following text. MP-$k$ measures algorithm's
performance on predicting most likely nodes.
\item[b)] Optimal $F_1$-score (referred later simply as $F_1$ Score). 
The optimal $F_1$-score is computed by computing $F_1$-scores ($\frac{2\times \text{precision} \times \text{recall}}{\text{precision} + \text{recall}}$)for each point on the Precision-Recall curve and selecting the largest $F_1$ value. Unlike MP-$k$, the optimal $F_1$ score is used to measure the
overall prediction performance of our algorithms. For instance, $F_1=0.5$ suggests the algorithm can reach a 
50\% precision at 50\% recall.
\end{itemize}

\subsection{Prediction using Composite Network}
To begin with, we illustrate different design aspects for our algorithm. 
%For each training process, we compute
%$s_u, u \in \{1,...,U\}$ as well as the optimal composite vector $(\alpha_1,...,\alpha_M)$ 
%using Eq. \ref{optimization}.
%During the testing process, for each app in the 
%testing set, we look at each user's neighbors,
%and compute the adoption likelihood using Eq. \ref{prob}.

To demonstrate the importance of modeling both networks and individual
variances in our model, 
we here demonstrate the prediction performance with five configurations using
a 5-fold cross-validation:
a) to model both individual variance and network effects; b) to model
both individual variance and network effects, but exclude the virtual 
network $G^p$ capturing exogenous factors;  c) to model 
with only individual variance (by forcing $\alpha_m =0$ in Eq. \ref{optimization}),
 d) to model with only network effects (by forcing $s_u=0, \forall u$), and
e) to model with network effects while allowing the composite vector to be negative.
The results are illustrated in Table \ref{comp1}.

\begin{table*}[ht]
\centering
\begin{tabular}{|c|c|c|c|}
\hline
& RMSE & MP-5 & $F_1$ Score \\
\hline
Net.+  Ind. Var. + Exogenous Factor  & {\bf 0.25} & {\bf 0.31} & {\bf 0.43} \\
\hline
Net. + Ind. Var. & 0.26  & 0.29 & 0.42 \\ 
\hline
Ind. Variance Only & 0.29 & 0.097 & 0.24 \\
\hline
Net. Only (non-negative) & 0.26 & 0.24 & 0.37 \\
\hline
Net. Only (allow negative) & 0.30 & 0.12 & 0.12 \\
\hline
\end{tabular}
\caption{The performance of our approach under five different configurations. We observe
that modeling both individual variance and networks are crucial in performance as well as
enforcing non-negative composition for candidate networks as in Eq. \ref{optg}.}
\label{comp1}
\end{table*}

We find the surprising results that app installations are highly predictable with individual 
variance and network information as shown in Table \ref{comp1}. In addition,
Table \ref{comp1} clearly suggests that all our assumptions for the 
model are indeed correct, and both individual variance and network effects play
important roles in app installation mechanism, as well as the exogenous factors modeled
by $G^p$.

We also notice that while accuracy almost doubles, it is often impossible
to realize this improvement using RMSE. Therefore, we will not RMSE for the rest of
the work. 

 \begin{figure}[h]
\centering
  \includegraphics[width=0.40\textwidth]{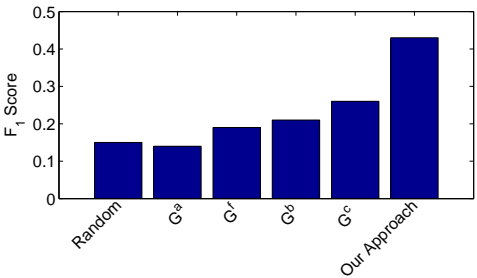}
\caption{We demonstrate the prediction performances using 
each single network here. For comparison, we also show the result of 
random guess, and the result using our approach, which combines all potential evidence. }
\label{change}
\end{figure}

We now illustrate the prediction performance when our algorithm is only
allowed to use one single network. The results are shown in Fig. \ref{change}. 
We find that except the affiliation network, almost all other networks predict
well above chance level. Bluetooth network performs much better than 
friendship networks, which matches previous work well\cite{eagle2009inferring}.
The call log network seems to achieve the best results,
which may due to the fact that we only take calls in the study community into
consideration. 
We conclude that while network effects are strong in app installations, a well-crafted
model such as our approach can vastly increase the performance by computing
the composite network and counting other factors in.

\subsection{Prediction Performance}
We now test the performance of our model with some other implementations
for predictions. As there is no other closer work related to
 app prediction with multiple networks, we here compare prediction performance 
with some alternative approaches we can think of.

Since it is practically difficult to observe every user app installation behaviors, in our
experiments we also want to test the performance of each algorithm 
when the test set is small. In particular, we 
evaluate the performance of different implementations with two approaches for cross validation:
1) Normal-size training set: We randomly choose half of all the apps in the dataset as the 
training set, and test on the other half of the dataset. 2) Small-size training set: We 
randomly choose only 20\% of all the app installations in our dataset as the training
set, and test on the the rest 80\% apps. In both cases, we repeat the process for five times
for cross validation and take average of the results.  

For our algorithm, we feed it with networks $G^p, G^a, G^b, G^f$ and 
$G^c$ obtained by phones and surveys as described previously. 
For SVM, we apply two different approaches in predictions: 
\begin{itemize}
\item We don't
consider the underlying network, but simply use the adoption status of all other nodes
as the features for each node. We test this approach simply to establish
a baseline for prediction. We refer it as ``SVM-raw''.
\item We compute the potential $p_a^m(i)$ for each candidate network $G^m$, and
we use all the potentials from all candidate networks as features. Therefore, 
we \emph{partially borrow} some ideas from our own model to implement this SVM approach. 
We refer this approach as ``SVM-hybrid''. 
\end{itemize}
We use a modern SVM implementation ~\cite{chih2001libsvm}, which is capable of generating 
probabilistic predictions rather than binary predictions. 

We also replace Eq. \ref{prob} with a linear regression model by using $p_a^m(i), \forall m$ 
together with \# of apps per user (instead of learning $s_u$ in our MLE framework) as independent variables.
We call this approach ``Our Approach (Regression)'' in the following text to distinguish the difference.
We also force the non-negative accumulation assumption in the regression setting.
%as it
%inherits many ideas from our model such as modeling individual variance,
%accumulative potentials and modeling exogenous factors.

\begin{table*}
\centering
\begin{tabular}{|c|cc|cc|cc|}
\hline
\multicolumn{1}{|c|}{\bf Methods} & \multicolumn{2}{|l|}{\bf Using 20\% as Training Set} & \multicolumn{2}{|l|}{\bf Using 50\% as Training Set} & \multicolumn{2}{|l|}{\bf Using 50\% as Training Set} 
\\ 
\multicolumn{1}{|c|}{} & \multicolumn{2}{|l|}{All Users} & \multicolumn{2}{|l|}{All Users} & \multicolumn{2}{|l|}{Low Activity Users} 
\\ \hline
 &  MP-$5$  &$F_1$ Score & MP-$5$ &$F_1$ Score & MP-$5$ & $F_1$ Score
\\ \hline 
Our Approach &{\bf 0.28} &{\bf 0.46} & {\bf 0.31}& {\bf 0.43} &{\bf 0.20} & {\bf 0.43}  \\ \hline
SVM-raw       &0.17&0.26&0.24&0.32&0.14&0.27\\ \hline
SVM-hybrid    &0.14&0.29&0.27&0.30&0.16&0.30\\  \hline
Our Approach (Regression) &0.27&0.42&0.30&0.41&0.18&0.39\\ \hline
%Regression-hybrid &0.24&0.40&0.23&0.40&0&0  \\ \hline
Random Guess  &0.081&0.17&0.081&0.17&0.076 &0.14
\\ \hline
\end{tabular}
\caption{Prediction performance for our algorithm and competing 
methods is shown.}
\label{large}
\end{table*}

Results for both the normal-size training set and the small-size training set
are shown in Table ~\ref{large}, and we discover that our algorithm outperforms other
competing approaches in all categories. However, we notice that with many our
model assumptions, generic methods can also achieve reasonably well results.
Performance on half of the users that are less active in app installation is also shown.
Because this group of users are very inactive, they may be more susceptible to 
network influence in app installation behaviors. We notice that 
our algorithm performs better in this group with more than 10\% improvement over other methods.

\subsection{Predicting Future Installations}
In app marketing, one key issue is to predict future app installations.
Predicting future app adoption at time $t$ in our model is equivalent to predicting 
installation with part of the neighbor adoption status unknown. These 
unknown neighbors who haven't adopted at time $t$ may or may not
adopt at $t'>t$.
Though our algorithm is trained without the information of time of adoption, we show here
that the inferred individual variance $s_u$ and composite vector $\left(\alpha_1,...,\alpha_M\right)$ can
be used to predict future app adoption. 

We here apply the following cross-validation scheme to test our 
algorithm's ability in predicting future installations:
For the adopters of each app, we split them to 
two equal-size groups by their time of adoption. Those who adopted earlier are
in G1, and those who adopted later are in G2. The training phase
is the same as the previous section; In the testing phase, each algorithm
will only see adoption information for subjects in G1, and predict
node adoption for the rest. The nodes in G2 will be marked as
non-adopters during prediction phase. 

Results from cross validation are shown in Table \ref{future}. We notice that our 
algorithm still maintains the best performance and limited decrease
in accuracy compared with Table \ref{large}. Since the number of adopted nodes
are fewer than those in Table \ref{large}, we here show MP with smaller $k$
in Table \ref{future}.

%It should be noted that
%the random guess base precision is reduced by half. Assume among 55 users,
%four of them adopt an app. If we predict with complete information, the 
%random guess precision is $\frac{4}{55}$. However, when we predict 
%future installation according to our scheme, the random guess precision
%is $\frac{2}{55-2}$, almost cut in half. 

\begin{table}[ht]
\begin{tabular}{|c|c c c|c|}
\hline 
 & \multicolumn{3}{|c|}{MP-$k$} & $F_1$ Score  \\
& \multicolumn{1}{|c}{$k=3$} & \multicolumn{1}{c}{$k=4$} & \multicolumn{1}{c|}{$k=5$} & \\
\hline 
Our Approach & {\bf 0.18} &{\bf 0.16}& {\bf 0.15}& {\bf 0.35} \\
\hline
SVM-hybrid & 0.15 & 0.13 & 0.12 & 0.32 \\
\hline
Our(Regression) & 0.17 & 0.15 & 0.14 & 0.33 \\
\hline
Random & 0.045 & 0.045 & 0.045 & 0.090 \\
\hline  
\end{tabular}
\caption{MP-$k$ and $F_1$ scores for predicting future app installations are shown above.}
\label{future}
\end{table}

Notice in Table ~\ref{future} that the random guess precision
is reduced by half. Therefore, even the precision here is 30\%
lower than in Table ~\ref{large}, it is mainly due to the fact
that nodes in G1 are no longer in the predicting set. Our accuracy
is considerable as it is four times better than random guess.

\subsection{Predictions With Missing Historical Data}
In practice, sometimes it is not possible to observe the app installation 
for all users due to privacy reasons. Instead,
for app market markers they may only be allowed to observe 
and instrument a small subset of a community.
We here want to study if it is still possible to make some prediction in app installations
under such circumstance.

\begin{figure}
\centering
\includegraphics[width=0.45\textwidth]{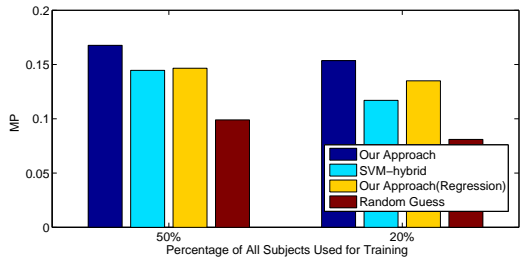}
\caption{The MP from our approach and two comparison approaches. We here set $k$ for MP to be the
average number of users in G2 for each testing app.} 
\label{incomplete}
\end{figure}

To formally state this problem, we assume that all the nodes $1,...,U$ are divided
into two groups. The observable group G1 and the unobservable group G2. During cross validation,
only nodes in the observable group are accessible to our algorithms in the training process, and
nodes in the unobservable group are tested with the prediction algorithms. Therefore, for
our algorithm, even the individual variance $s_u, u \in \text{G1}$ is computed in the training process, 
we will not have $s_{u'}, u' \in \text{G2}$ for Eq. \ref{prob} in the testing phase. 
We illustrate the prediction precision results in Fig. ~\ref{incomplete}.
It seems that even trained on a different
set of subjects without calibrating users variance, the composite vector learned by our algorithm
 can still be applied to another set of users and achieve 80\% over random guess.

\section{Conclusion}
Our contributions in this paper include a) We show the data of a novel mobile phone
based experiments on the app installation behavior; b) We illustrate that
there are strong network effects in app installation patterns even with
tremendous uncertainty in app installation behavior; c) We show that by
combining measurable networks using modern smart phones, we can maximize
the prediction accuracy; d) We develop  
a simple discriminative model which combines individual variance, multiple networks
and exogneous factors, and our model provides prediction accuracy four times 
better than random guess in predicting future installations.

Future works include the causality problem in studying network phenomena and a temporal
model for app adoption. We believe the former one can be done with a 
much carefully crafted lab experiments. For the latter one, we have attempted 
multiple temporal adoption models but failed.We suspect that the mechanism of 
temporal diffusion of apps is very complicated, and we leave 
this as a future work.

Though our convex optimization framework is fast and reasonably scalable, 
it should be noted that still the proposed method in this paper 
may not be suitable to handle data from billions of cell phone users. 
Potential solutions include dividing users into small clusters and then conquering,
and sampling users for computation. The scalability problem remains a future
work.

\section{Acknowledgements}
The authors want to thank Cory Ip for her remarkable efforts in managing
the study, Dr. Riley Crane and Yves-Alexandre de Montjoye for helpful discussions,
and the anonymous reviewers for their valuable comments.
This research was sponsored by AFOSR under Award Number FA9550-10-1-0122.
The views and conclusions contained in this document are those of the
authors and should not be interpreted as representing the official
policies, either expressed or implied, of AFOSR or the U.S. Government.

{
\bibliographystyle{aaai}
\bibliography{panwei}
}

\end{document}